\documentclass[11pt]{article}
\usepackage[hmargin=3cm,vmargin=2cm]{geometry}
\usepackage{graphicx, amssymb}
\title{\bf Dirac Equation with Spin Symmetry for the Modified P\"oschl-Teller Potential in $D$-dimensions}
\author{\bf D. Agboola\footnote{e-mail:~tomdavids2k6@yahoo.com}}
\date{\it Department of Mathematics,College of Science and Technology,Covenant University, Ogun State, P.M.B.1023, Nigeria.}
\begin{document}
\maketitle
\vspace{0.5in}
\noindent {\bf Abstract} We present solutions of the Dirac equation with spin symmetry for vector and scalar modified P\"oschl-Teller potential within framework of an approximation of the centrifugal term. The relativistic energy spectrum is obtained using the Nikiforov-Uvarov method and the two-component spinor wavefunctions are obtain are in terms of the Jacobi polynomials. It is found that there exist only positive-energy states for bound states under spin symmetry, and the energy levels increase with the dimension and the potential range parameter $\alpha$.

\maketitle
\vspace{0.5in}
\vspace{.5in}


\maketitle
\vspace{1in}

Solutions to relativistic equations play a very important in many aspects of mordern physics. In particular, the Dirac equation which describe the motion of a spin-$\frac{1}{2}$ particle has been used in solving many problems of nuclear aand high-energy physics. The spin symmetry arises if the magnitude of the attractive Lorentz scalar potential $S(r)$ and the replusive vector potential $V(r)$ are nearly equal, i.e $S(r)\sim V(r)$, in the nuclei, while the pseudo-spin symmetry occur when $S(r)\sim -V(r)$ [1-3]. The case of the exact spin and pseudo-spin symmetries has been shown to correspond to the $SU(2)$ symmetries of the Dirac Hamiltonian [3]. The spin symmetry is relevant for mesons [4] and the pseudo-spin symmetry is used to explain deformed nuclei [5], super-deformation [6] and to establish an effective nuclear shell-model scheme [7-9]. Also, various potentials such as the Morse potential [10-12], Wood-saxon potential [13], Coulomb and Hartmann potentials [14], Eckart potential [15, 16], P\"oschl-Teller potential [17, 18] and the harmonic potential [19, 20] have been studied within the frame work of the spin and pseudospin symmetries.   

Moreover, with the interest in higher dimensional theory, the multi-dimensional quantum mechanical equations-relativistic and nonrelativistic- have been solved with various physical potentials. To mention a few, the $D$-dimensional Schr\"{o}dinger has been studied with the Coulomb-like potential [21], pseudoharmonic potential [22], Hulth$\acute{e}$n potential [23] and P\"oschl-Teller potential [24]. In addition, the $D$-dimensional relativistic Klein-Gordon and Dirac equations  have been studied with many exactly solvable models [25-30]. However, some physical potentials can only be solve exactly for the $s$-states. Unfortunately, the modified P\"oschl-Teller potential is one of these potentials. For instance, a recent work [31] has presented the $s$-wave solutions of the Dirac equation with the P\"{o}schl-Teller potential under the conditions of the exact spin symmetry and pseudospin symmetry. However, in order to extend the solutions of the modified P\"oschl-Teller potential to any $\ell\neq 0$ state, some recent studies [17, 24] have used a hyperbolic approximation for the centrifugal term to obtain the non-relativistic solutions of the modified P\"oschl-Teller potential. In light of this, the present Letter intends to extend the discussions on the relativistic P\"{o}schl-Teller potential to $D$-dimensions by presenting the bound-state solutions of the $D$-dimensions Dirac equation with spin symmetry for Lorentz vector and scalar modified P\"oschl-Teller potential using the Nikiforov-Uvarov method [32].

The $D$-dimensional Dirac equation with a scalar potential $V_s(r)$ and a vector potential $V_v(r)$ and mass $\mu$ can be written in natural units $\hbar=c=1$ as [27, 33, 34]
$$H\Psi{(r)}=E_{n_r\kappa}\Psi(r)\hspace{.2in}\mbox{where}\hspace{.1in}H=\sum_{j=1}^D\hat{\alpha}_jp_j+\hat{\beta}[\mu+V_s(r)]+V_v(r) \eqno{(1)}$$ where $E_{n_r\kappa}$ is the relativistic energy, $\{\hat{\alpha}_j\}$ and $\hat{\beta}$ are Dirac matrices, which satisfy anti-commutation relations
$$\begin{array}{lrl}
\hat{\alpha}_j\hat{\alpha}_k+\hat{\alpha}_k\hat{\alpha}_j&=&2\delta_{jk}\bf{1}\\
\hat{\alpha}_j\hat{\beta}+\hat{\beta}\hat{\alpha}_j&=&0\\
{\hat{\alpha}_j}^2=\hat{\beta}^2&=&\bf{1}

\end{array} \eqno{(2)}$$
and 
$$p_j=-i\partial_j=-i\frac{\partial}{\partial x_j} \hspace{.2in} 1\leqslant j\leqslant D. \eqno{(3)}$$ The orbital angular momentum operators $L_{jk}$, the spinor opertaors $S_{jk}$ and the total angular momentum operators $J_{jk}$ can be defined as follows:
$$L_{jk}=-L_{jk}=ix_j\frac{\partial}{\partial x_k}-ix_k\frac{\partial}{\partial x_j},\hspace{.2in} S_{jk}=-S_{kj}=i\hat{\alpha}_j\hat{\alpha}_k/2,\hspace{.2in} J_{jk}=L_{jk}+S_{jk}.$$
$$L^2=\sum_{j<k}^DL^2_{jk},\hspace{.2in}S^2=\sum_{j<k}^DS^2_{jk},\hspace{.2in}J^2=\sum_{j<k}^DJ^2_{jk}, \hspace{.2in} 1\leqslant j< k\leqslant D. \eqno{(4)}$$
For a spherically symmetric potential, total angular momentum operator $J_{jk}$ and the spin-orbit operator $\hat{K}=-\hat{\beta}(J^2-L^2-S^2+(D-1)/2)$ commutate with the Dirac Hamiltonian. For a given total angular momentum $j$, the eigenvalues of $\hat{K}$ are $\kappa=\pm(j+(D-2)/2)$; $\kappa=-(j+(D-2)/2)$ for aligned spin $j=\ell+\frac{1}{2}$ and $\kappa=(j+(D-2)/2)$ for unaligned spin $j=\ell-\frac{1}{2}$.

Thus, we can introduce the hyperspherical coordinates [35]
$$\begin{array}{lrl}
x_1&=&r\cos\theta_1\\
x_\alpha&=&r\sin\theta_1\dots\sin\theta_{\alpha-1}\cos\phi,\hspace{.2in}2\leqslant \alpha\leqslant D-1\\
x_D&=&r\sin\theta_1\dots\sin\theta_{D-2}\sin\phi,\\
\end{array} \eqno{(5)}$$ where the volume element of the configuration space is given as
$$\prod_{j=1}^Ddx_j=r^{D-1}drd\Omega \hspace{.2in} d\Omega=\prod_{j=1}^{D-1}(\sin\theta_j)^{j-1}d\theta_j  \eqno{(6)}$$
with $0\leqslant r< \infty$, \hspace{.1in}$0\leqslant\theta_k\leqslant\pi$, $k=1,2,\dots D-2$,\hspace{.1in} $0\leqslant\phi\leqslant 2\pi$, such that the spinor wavefunctions can be classified according to the hyperradial quantum number $n_r$ and the spin-orbit quantum number $\kappa$ and can be written using the Pauli-Dirac representation 
$$\Psi_{n_r\kappa}(r,\Omega_D)=r^{-\frac{D-1}{2}}\left(\begin{array}{lll}
F_{n_r\kappa}(r)Y_{jm}^\ell\left(\Omega_{D}\right)\\\\
iG_{n_r\kappa}(r)Y^{\tilde{\ell}}_{jm}\left(\Omega_{D}\right)
\end{array}\right) \eqno{(7)}$$ 
where $F_{n_r\kappa}(r)$ and $G_{n_r\kappa}(r)$ are the radial wave function of the upper- and the lower-spinor components respectively, $Y_{jm}^\ell\left(\Omega_{D}\right)$ and $Y^{\tilde{\ell}}_{jm}\left(\Omega_{D}\right)$ are the hyperspherical harmonic functions coupled with the total angular momentum $j$. The orbital and the pseudo-orbital angular momentum quantum numbers for spin symmetry  $\ell$ and  and pseudospin symmetry $\tilde{\ell}$ refer to the upper- and lower-component respectively.  

Substituting Eq.\,(7) into Eq.\,(1), and seperating the variables we obtain the following coupled radial Dirac equation for the spinor components:
$$\left(\frac{d}{dr}+\frac{\kappa}{r}\right)F_{n_r\kappa}(r)=[\mu+E_{n_r\kappa}-\Delta(r)]G_{n_r\kappa}(r)\eqno{(8)}$$
$$\left(\frac{d}{dr}-\frac{\kappa}{r}\right)G_{n_r\kappa}(r)=[\mu-E_{n_r\kappa}+\Sigma(r)]F_{n_r\kappa}(r)\eqno{(9)}$$
where $\Delta(r)=V_v(r)-V_s(r)$, $\Sigma(r)=V_v(r)+V_s(r)$ and  $\kappa=\pm(2\ell+D-1)/2$. Further details of the derivation can be obtain from refs [36-38]. Using Eq.\,(8) as the upper component and substituting into Eq.\,(9), we obtain the follwoing second order differential equations
$$\left[\frac{d^2}{dr^2}-\frac{\kappa(\kappa+1)}{r^2}-[\mu+E_{n_r\kappa}-\Delta(r)][\mu-E_{n_r\kappa}+\Sigma(r)]+\frac{\frac{d\Delta(r)}{dr}\left(\frac{d}{dr}+\frac{\kappa}{r}\right)}{[\mu(r)+E_{n_r\kappa}-\Delta(r)]}\right]F_{n_r\kappa}(r)=0 \eqno{(10)}$$
$$\left[\frac{d^2}{dr^2}-\frac{\kappa(\kappa-1)}{r^2}-[\mu+E_{n_r\kappa}-\Delta(r)][\mu-E_{n_r\kappa}+\Sigma(r)]-\frac{\frac{d\Sigma(r)}{dr}\left(\frac{d}{dr}-\frac{\kappa}{r}\right)}{[\mu(r)-E_{n_r\kappa}+\Sigma(r)]}\right]G_{n_r\kappa}(r)=0 \eqno{(11)}$$
We note that the energy eigenvalues in these equation depend on the angular momentum quantum number $\ell$ and dimension $D$. However, to solve these equations, we shall use an approximation for the centrifugal barrier and obtain the solutions using the Nikiforov-Uvarov method.

Next, we give a brief description of the conventional Nikiforov-Uvarov method. A more detailed description of the method can be obtained the following reference [32]. With an appropriate transformation $s=s(r)$,the one dimensional Schr$\ddot{o}$dinger equation can be reduced to a generalized equation of hypergeometric type which can be written as follows:
$$\psi ''(s)+ \frac{\tilde{\tau}(s)}{\sigma(s)}\psi '(s)+ \frac{\tilde{\sigma}(s)}{\sigma^2(s)}\psi(s)=0  \eqno{(12)}$$ 
Where $\sigma(s)$ and $\tilde{\sigma}(s)$ are polynomials, at most second-degree, and $\tilde{\tau}(s)$is at most a first-order polynomial. To find particular solution of Eq.\,(12) by separation of variables, if one deals with
$$\psi(s)=\phi(s)y_{n_r}(s),  \eqno{(13)}$$
Eq.\,(12) becomes
$$\sigma(s)y ''_{n_r}+\tau(s)y '_{n_r} +\lambda y_{n_r} =0  \eqno{(14)}$$
where
$$\sigma(s)= \pi(s)\frac{\phi(s)}{\phi '(s)}  \eqno{(15)}$$
$$\tau(s)=\tilde{\tau}(s)+2\pi(s) ,  \tau '(s)<0,  \eqno{(16)}$$\\
$$\pi(s)=\frac{\sigma^\prime-\tilde{\tau}}{2}\pm \sqrt{\left(\frac{\sigma '-\tilde{\tau}}{2}\right)^2-\tilde{\sigma}+t\sigma},  \eqno{(17)}$$
and 
$$\lambda=t+\pi '(s).  \eqno{(18)}$$
The polynomial $\tau(s)$ with the parameter $s$ and prime factors show the differentials at first degree be negative. However, determination of parameter $t$ is the essential point in the calculation of $\pi(s)$. It is simply defined by setting the discriminate of the square root to zero [32]. Therefore, one gets a general quadratic equation for $t$. The values of $t$ can be used for calculation of energy eigenvalues using the following equation
$$\lambda=t+\pi '(s)=-n_r\tau '(s)-\frac{n_r(n_r-1)}{2}\sigma ''(s).   \eqno{(19)}$$
Furthermore, the other part $y_{n_r}(s)$ of the wave function in Eq. (12) is the hypergeometric-type function whose polynomial solutions are given by Rodrigues relation: 
$$y_{n_r}(s)=\frac{B_{n_r}}{\rho(s)}\frac{d^{n_r}}{ds^{n_r}}[\sigma^{n_r}(s)\rho(s)]  \eqno{(20)}$$ 
where $B_{n_r}$ is a normalizing constant and the weight function $\rho(s)$ must satisfy the condition [32]
$$(\sigma\rho) ' =\tau\rho.   \eqno{(21)}$$

The Lorentz vector $V_v(r)$ and scalar $V_s(r)$ modified P\"oschl-Teller potential can be defined as follows [24, 39-41]
$$V_v(r)=-\frac{V_0}{\cosh^2(\alpha r)}\hspace{.2in} \mbox{and}\hspace{.2in} V_s(r)=-\frac{S_0}{\cosh^2(\alpha r)}   \eqno{(22)}$$ where $\alpha$ is related to the range of the potential and $V_0$ and $S_0$ are the depths of the vector and scalar potentials respectively. Moreover, we can approximate the centrifugal terms as follows [17, 24]
 $$\frac{1}{r^2}\approx\frac{\alpha^2}{\sinh^2(\alpha r)}.\eqno(23)$$ Substituting Eqs.\,(22) and (23) into Eqs.\,(10) and (11), we have 
$$\left[\frac{d^2}{dr^2}-\frac{\alpha^2\kappa(\kappa+1)}{\sinh^2(\alpha r)}-[\mu+E_{n_r\kappa}-\Delta(r)][\mu-E_{n_r\kappa}+\Sigma(r)]+\frac{\frac{d\Delta(r)}{dr}\left(\frac{d}{dr}+\frac{\kappa}{r}\right)}{[\mu(r)+E_{n_r\kappa}-\Delta(r)]}\right]F_{n_r\kappa}(r)=0 \eqno{(24)}$$
$$\left[\frac{d^2}{dr^2}-\frac{\alpha^2\kappa(\kappa-1)}{\sinh^2(\alpha r)}-[\mu+E_{n_r\kappa}-\Delta(r)][\mu-E_{n_r\kappa}+\Sigma(r)]-\frac{\frac{d\Sigma(r)}{dr}\left(\frac{d}{dr}-\frac{\kappa}{r}\right)}{[\mu(r)-E_{n_r\kappa}+\Sigma(r)]}\right]G_{n_r\kappa}(r)=0 \eqno{(25)}$$ 
where 
$$\Delta(r)=\frac{S_0-V_0}{\cosh^2(\alpha r)}\hspace{.2in} \mbox{and}\hspace{.2in} \Sigma(r)=\frac{-(V_0+S_0)}{\cosh^2(\alpha r)}. \eqno{(26)}$$
For the case of spin symmetry, $V_v(r)\sim V_s(r)$, i.e. $\Delta(r)=V_v(r)-V_s(r)=C_1$ (a constant), which implies that $\frac{d\Delta(r)}{dr}=0$. Thus, putting this into Eq.\,(24), we have
$$\left[\frac{d^2}{dr^2}-\frac{\alpha^2\kappa(\kappa+1)}{\sinh^2(\alpha r)}-(\mu-E_{n_r\kappa})(\mu+E_{n_r\kappa}-C_1)+\frac{(V_0+S_0)(E_{n_r\kappa}+\mu-C_1)}{\cosh^2(\alpha r)}\right]F_{n_r\kappa}(r)=0. \eqno{(27)}$$ 
If we take the transformation $s=\tanh^2(\alpha r)$, Eq.\,(27) becomes 
$$F''_{n_r\kappa}(s)+\frac{1-3s}{2s(1-s)}F'_{n_r\kappa}(s)+\frac{1}{4s^2(1-s)^2}[-\delta s^2+(\delta+\gamma-\epsilon^2)s-\gamma]F_{n_r\kappa}=0 \eqno{(28)}$$ 
where $$ \epsilon^2=\frac{(\mu-E_{n_r\kappa})(\mu+E_{n_r\kappa}-C_1)}{\alpha^2},\hspace{.2in} \delta=\frac{(V_0+S_0)(E_{n_r\kappa}+\mu-C_1)}{\alpha^2}\hspace{.1in}  \mbox{and}\hspace{.1in}  \gamma=\kappa(\kappa+1).  \eqno{(29)}$$ 
Comparing Eqs.\,(28) and (12) we can define the following
$$\tilde{\tau}(s)=1-3s,\hspace{.1in}\sigma(s)=2s(1-s)\hspace{.1in} \mbox{and}\hspace{.1in}\tilde{\sigma}(s)=-\delta s^2+(\gamma+\delta-\epsilon^2)s-\gamma \eqno{(30)}$$
Inserting these into Eq.\,(17), we have the following function $$\pi(s)=\frac{1-s}{2}\pm\frac{1}{2}\sqrt{(1+4\delta-8t)s^2+(8t-4(\gamma+\delta-\epsilon^2)-2)s+4\gamma+1}  \eqno{(31)}$$ The constant parameter $t$ can be found by the condition that the discriminant of the expression under the square root  has a double root, i.e., its discriminant is zero. Thus the possible value function for each value of $t$ is given as 
$$\pi(s)=\frac{1-s}{2}\pm \left\{\begin{array}{lll}
\frac {1}{2}\left [\left(-2\epsilon +\sqrt{1+4\gamma}\right)s-\sqrt{1+4\gamma}\right] & \mbox{for} & t=-\frac{1}{2}(\gamma-\delta+\epsilon^2)+\frac{1}{2}\epsilon\sqrt{1+4\gamma}\\\\
\frac {1}{2}\left[\left(2\epsilon +\sqrt{1+4\gamma}\right)s-\sqrt{1+4\gamma}\right] & \mbox{for} & t=-\frac{1}{2}(\gamma-\delta+\epsilon^2)-\frac{1}{2}\epsilon\sqrt{1+4\gamma}\end{array}\right. \eqno{(32)}$$\\
By Nikiforov-Uvarov method, we made an appropriate choice of the function $\pi(s)=\frac{1-s}{2}-\frac {1}{2}\left[\left(2\epsilon +\sqrt{1+4\gamma}\right)s-\sqrt{1+4\gamma}\right]$ such that by Eq.\,(19), we can obtain the eigenvalue equation to be $$-\frac{1}{2}(\gamma-\delta+\epsilon^2)-\frac{1}{2}\epsilon\sqrt{1+4\gamma}-\frac{1}{2}(2\epsilon+\sqrt{1+4\gamma})-\frac{1}{2}=n_r[4+2\epsilon+\sqrt{1+4\gamma}]+2n_r(n_r-1)  \eqno{(33)}$$ 
Eq.\,(33) can be written in the powers of $\epsilon$ as follows
$$\epsilon^2+\epsilon\left[2(2n_r+1)+\sqrt{1+4\gamma}\right]+(\gamma-\delta)+\left[(1+2n_r)+\sqrt{1+4\gamma}\right]=0,  \eqno{(34)}$$	
such that we can obtain
$$-\epsilon^2=-\frac{1}{4}\left[-2(2n_r+1)-\sqrt{1+4\gamma}+\sqrt{1+4\delta}\right]^2, \eqno{(35)}$$
from which we can obtain a rather complicated transcendental
energy equation:
$$(\mu-E_{n_r\kappa})(\mu+E_{n_r\kappa}-C_1)=\frac{\alpha^2}{4}\left[2(2n_r+1)+(2\kappa+1)-\frac{1}{\alpha}\sqrt{\alpha^2+4(V_0+S_0)(E_{n_r\kappa}+\mu-C_1)}\right]^2 \eqno{(36)}$$
If we define a principal quantum number $n=2n_r+\ell+1$ , Eq.\,(36) becomes
$$(\mu-E_n)(\mu+E_n-C_1)=\frac{\alpha^2}{4}\left[2n+D-\frac{1}{\alpha}\sqrt{\alpha^2+4(V_0+S_0)(E_n+\mu-C_1)}\right]^2 \eqno{(37)}$$ where we have chosen $\kappa=(2\ell+D-1)/2$ and $n=1,2,3,\dots$ Some numerical values of the energy levels $E(\alpha,n,D)$ for some dimensions and exited states are given in Table 1.

We now obtain the spinor components of the  wavefunction for the spin symmetry case using the Nikiforov-Uvarov method. By substituting $\pi(s)$ and $\sigma(s)$ into Eq.\,(15), and solving the first order differential  equation to have
$$\phi(s)=s^{(\kappa+1)/2}(1-s)^{\epsilon/2}.  \eqno{(38)}$$
Also using Eq.\,(18), the weight function $\rho(s)$ can be obtained as 
$$\rho (s)=\frac{1}{2}s^{(2\kappa-1)/2}(1-s)^\epsilon  \eqno{(39)}$$
Substituting Eq.\,(39) into the Rodrigues relation (20), we have
$$y_{n_r}(s)=B_{n_r}s^{-(2\kappa-1)/2}(1-s)^{-\epsilon}\frac{d^{n_r}}{ds^{n_r}}\left[s^{n_r+(2\kappa-1)/2}(1-s)^{n_r+\epsilon}\right]. \eqno{(40)}$$ 
Therefore, we can write the upper component $F_{{n_r}\kappa}(s)$ as 
$$F_{{n_r}\kappa}(s)=C_{n_r}s^{(\kappa+1)/2}(1-s)^{\epsilon/2}P^{((2\kappa-1)/2,~\epsilon)}_{n_r}(1-2s)  \eqno{(41)}$$
where $C_{n_r}$ is the normalization constant, and we have used the definition of the Jacobi polynomials [42], given as
$$P^{(a,~b)}_n(s)=\frac{(-1)^n}{n!2^n(1-s)^a(1+s)^b}\frac{d^n}{ds^n}\left[(1-s)^{a+n}(1+s)^{b+n}\right].  \eqno{(42)}$$ The lower-component can be obtain as follows using Eq.\,(8)
$$G_{n_r\kappa}(s)=A_1(s)P_{n_r}^{((2\kappa-1)/2,~\epsilon)}(1-2s)+ A_2(s)P_{n_r-1}^{(2\kappa+1)/2,~\epsilon/2+1)}(1-2s) \eqno{(43)}$$ where
$$A_1(s)=\frac{C_{n_r}\alpha s^{\kappa/2}(1-s)^{\epsilon/2}\left[\left(\frac{\kappa+1}{2}\right)(1-s)-\frac{\epsilon}{2}s\right]+C_{n_r}\frac{\alpha\kappa}{\tanh^{-1}(\sqrt{s})}}{\mu+E_{n_r\kappa}-C_1}$$ and 
$$\hspace{.1in} A_2(s)=\frac{D_{n_r}\alpha s^{(\kappa+2)/2}(1-s)^{(\epsilon+2)/2}}{\mu+E_{n_r\kappa}-C_1} \eqno{(44)}$$ with constant $D_{n_r}$ defined by
$$D_{n_r}=\frac{2\epsilon+2n_r+2\kappa+1}{4}\times C_{n_r} \eqno{(45)}$$.
Moreover, to compute the normalization constant $C_{n_r}$, it is easy to show that $$\int^\infty_0\left|r^{\frac{-(D-1)}{2}}F_{{n_r}\kappa}(r)\right|^2r^{D-1}dr=\int^\infty_0|F_{{n_r}\kappa}(r)|^2dr=\int^1_0|F_{{n_r}\kappa}(s)|^2\frac{ds}{2\alpha\sqrt{s}(1-s)}=1  \eqno{(46)}$$
where we have also used the substitution $s=\tanh^2(\alpha r)$. Putting Eq.\,(41) into Eq.\,(46) and using the following definition of the Jacobi polynomial [42]
$$P^{(a,b)}_n(s)=\frac{\Gamma(n+a+1)}{n!\Gamma(1+a)} \ _2F_1\left(-n,a+b+n+1;1+a;\frac{1-s}{2}\right),  \eqno{(47)}$$ we arrived at 
$$C_{n_r}^2 N_{n_r}\int_0^1s^{\kappa+\frac{1}{2}}(1-s)^{\epsilon-1}[\ _2F_1\left(-n_r,\kappa+\epsilon+n_r+1/2;~\kappa+1/2;~s\right)]^2ds=\alpha \eqno{(48)}$$ where $N_{n_r}=\frac{1}{2}\left[\frac{\Gamma(n_r+\kappa+1/2)}{n_r!\Gamma(\kappa+1/2)}\right]^2$ and $_2F_1$ is the hypergeometric function. Using the following series representation of the hypergeometric function
$$_pF_q(a_1,...,a_p;c_1,...,c_q;s)=\sum_{n=0}^\infty\frac{(a_1)_n...(a_p)_n}{(c_1)_n...(c_q)_n}\frac{s^n}{n!}  \eqno{(49)}$$
we have
$$ C_{n_r}^2 N_{n_r}\sum^{n_r}_{i=0}\sum^{n_r}_{j=0}\frac{(-n_r)_i(\kappa+\epsilon+n_r+1/2)_i}{(\kappa+1/2)_i i!}\frac{(-n_r)_j(\kappa+\epsilon+n_r+1/2)_j}{(\kappa+1/2)_j j!}\int_0^1s^{\kappa+i+j+\frac{1}{2}}(1-s)^{\epsilon-1} ds=\alpha .\eqno{(50)}$$
Hence, by the definition of the Beta function, Eq.\,(43) becomes
$$\small C_{n_r}^2 N_{n_r}\sum^{n_r}_{i=0}\sum^{n_r}_{j=0}\frac{(-n_r)_i(\kappa+\epsilon+n_r+1/2)_i}{\kappa+1/2)_i i!}\frac{(-n_r)_j(\kappa+\epsilon+n_r+1/2)_j}{(\kappa+1/2)_j j!} B\left(\kappa+i+j+\frac{3}{2},\epsilon\right)=\alpha.  \eqno{(51)}$$
Using the relations $B(x,y)=\frac{\Gamma(x)\Gamma(y)}{\Gamma(x+y)}$ and the Pochhammer symbol $(a)_n=\frac{\Gamma(a+n)}{\Gamma(a)}$, Eq.\,(51) can be written as \\
$$\small C_{n_r}^2 N_{n_r}\sum^{n_r}_{i=0}\frac{(-n_r)_k(\kappa+\epsilon+n_r+1/2)_i(\kappa+\frac{3}{2})_i}{(\epsilon+\kappa+\frac{3}{2})_k (\kappa+1/2)_i k!}\sum^{n_r}_{j=0}\frac{(-n_r)_j(\kappa+\epsilon+n_r+1/2)_j(\kappa+i+\frac{3}{2})_j}{(\epsilon+\kappa+i+\frac{3}{2})_j (\kappa+1/2)_j j!}=\frac{\alpha}{B(\kappa+\frac{3}{2},\epsilon)}  \eqno{(52)}$$

\noindent Lastly, Eq.\,(52) can be used to compute the normalization constants for $n_r=0,1,2,...$ In particular for the ground state, i.e $n_r=0$, we have
$$C_0=\sqrt{\frac{2\alpha}{B\left(\kappa+\frac{3}{2},\epsilon\right)}}.  \eqno{(53)}$$


In conclusion, the solutions of the Dirac equation with spin symmetry for the modified P\"oschl-Teller potential has been extended to a multi-dimensional case. The energy levels and the spinor-components of the wavefunction were obtained using the Nikiforov-Uvarov method. We also obtain the normalization constants in form of the hypergeometric series. Numerical results show that there are only positive-energy states for bound states with spin symmetry. Also, the energy levels increase with the dimension and the potential range parameter $\alpha$. Moreover, the existence of the degenerate states between $E(\alpha,n+1,D)$ and $E(\alpha,n,D+2)$ indicate that the energy levels can be completely determined using the ground state.

\begin{table}[h]
\caption{ The bound-state energy levels $E_n$ are shown in the case of spin symmetry. The numerical results show that the energy levels increase with  both the dimensions $D$ and the range parameter $\alpha$.}

\centering
\begin{tabular}{|l|l|c|c|c|c|}\hline
\multicolumn{6}{|c|}{$E(\alpha,n,D)$}\\\hline
\multicolumn{6}{|c|}{$C_1=V_0=S_0=\mu=1$}\\\hline
$D$&$n$&$\alpha=0.0001$&{$\alpha=0.001$}&$\alpha=0.005$&$\alpha=0.01$\\\hline
&1&4.0032$\times 10^{-8}$&4.0326$\times 10^{-6}$&1.0442$\times 10^{-4}$&4.4016$\times 10^{-4}$\\
&2&9.0108$\times 10^{-8}$&9.1118$\times 10^{-6}$&2.4161$\times 10^{-4}$&1.1130$\times 10^{-3}$\\
3&3&1.6026$\times 10^{-7}$&1.6271$\times 10^{-5}$&4.4842$\times 10^{-4}$&---\\
&4&2.5050$\times 10^{-7}$&2.5544$\times 10^{-5}$&---&---\\
&5&3.6087$\times 10^{-7}$&3.6973$\times 10^{-5}$&---&---\\\hline
&1&6.2562$\times 10^{-8}$&6.3142$\times 10^{-6}$&1.6530$\times 10^{-4}$&7.1490$\times 10^{-4}$\\
&2&1.2267$\times 10^{-7}$&1.2429$\times 10^{-5}$&3.3488$\times 10^{-4}$&---\\
4&3&2.0287$\times 10^{-7}$&2.0641$\times 10^{-5}$&6.0121$\times 10^{-4}$&---\\
&4&3.0317$\times 10^{-7}$&3.0955$\times 10^{-5}$&---&---\\
&5&4.2361$\times 10^{-7}$&4.3513$\times 10^{-5}$&---&---\\\hline
&1&9.0108$\times 10^{-8}$&9.1118$\times 10^{-6}$&2.4161$\times 10^{-4}$&1.1131$\times 10^{-3}$\\
&2&1.6026$\times 10^{-7}$&1.6270$\times 10^{-5}$&4.4842$\times 10^{-4}$&---\\
5&3&2.5050$\times 10^{-7}$&2.5541$\times 10^{-5}$&---&---\\
&4&3.6087$\times 10^{-7}$&3.6973$\times 10^{-5}$&---&---\\
&5&4.9139$\times 10^{-7}$&5.0616$\times 10^{-5}$&---&---\\\hline
\end{tabular}

\label{tab:}
\end{table}

\end{document}